# Validation of a new 60 MeV proton beam-line for radiation hardness testing

Petter Hofverberg[1], Francoise Bezerra[2], Marine Ruffenach[2], Joêl Hérault[1], Julien Mekki[2], Arnaud Dufour[2], Richard Trimaud[1]
[1]*Centre Antoine Lacassagne, Nice, France*
[2]*CNES, Toulouse, France*

*Abstract*—A 60 MeV proton beam-line has been developed in Nice, France, in collaboration with the *Centre National d'Etudes Spatiales* (CNES). Experimental results are presented here to validate the beam-line for radiation hardness testing.

## I  INTRODUCTION

A new proton beam-line dedicated to R&D programs has recently been developed at *Centre Antoine Lacassagne* (CAL), in Nice (France), in collaboration with the *Centre National d'Etudes Spatiales* (CNES) [1]. This is the second beam-line of the MEDICYC 65 MeV cyclotron that is currently in operation, the first being the clinical 'eye-line' used for ocular proton therapy [2,3]. The R&D beam-line is proposed with two configurations, the first producing a Gaussian narrow beam of a few mm width, the second a 100 mm diameter flat beam. The second beam configuration has been developed in particular with irradiation tolerance tests of electronic components in mind. This facility could be used for various tests such as calibration of detectors, total non ionizing dose deposition or single event testing for components sensitive below 60 MeV. Following the development and commissioning of the R&D beam-line in 2020 and 2021 by CAL, a campaign has been conducted by CNES to validate the performance of the beam-line for radiation tolerance tests of electronic devices or systems. This paper presents a brief summary of the beam-line commissioning results and preliminary results from the validation campaign by CNES.

## II  THE IRRADIATION FACILITY

The R&D beam-line splits from the clinical beam-line at the second dipole magnet, M4, shown in Fig 1. A beam nozzle has been installed shortly after the dipole and contains the necessary instrumentation to monitor the beam and to control the beam delivery. A precision irradiation table is mounted in the prolongation of the beam and allows beam instrumentation and irradiation targets to be positioned accurately with respect to the beam axis (Fig 2). Operators and researchers can enter the irradiation room through a ~25 m long concrete block labyrinth. Access to the labyrinth is protected by a locked door which can be opened only when the activation in the irradiation room falls below a given threshold,. A control room is located on the far side of the labyrinth and provides the control systems necessary to perform beam calibration and beam delivery. The control room also contains all normal office equipment for guest researchers to perform their work.

Standard cables (BNC, SHV, RJ45 and fiber) are preinstalled between the irradiation room and the control room and are freely available for users.

The device-under-test (DUT) is positioned in the beam-line by means of a fixation frame, originally designed by UCL [4] . Custom fixation frames can also be made on demand. The frame is positioned on the irradiation table at a distance of 229 cm from the beam nozzle, 5 cm downstream of the final 100 mm collimator. The collimator frame also contains the PMMA range shifters that can optionally be added to obtain a specific energy at the DUT, down to 20 MeV. Read-out electronics can either be positioned in the control room, in the labyrinth or next to the irradiation table depending on its tolerance to radiation. The required cable length in each case is 25 m, 8 m and 3 m respectively. The radiation inside the labyrinth is for most applications negligible, but the area is however not accessible for personnel during an irradiation. Polyethylene or lead shielding can be provided for equipment positioned next to the irradiation table, but sensitive equipment should be placed elsewhere.

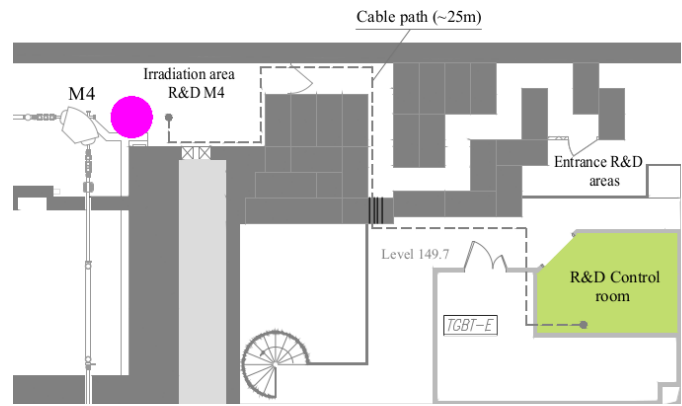

*Figure 1: The R&D irradiation room (top left), the control room (lower right), and the labyrinth through which the irradiation room can be accessed.*

## III  COMMISSIONING RESULTS

### III.A  Beam flux

The stability of the extracted beam proton current between 10 pA and 10 uA has been measured to be better than 0.1% above 100 pA and 1% below 100 pA. This beam current corresponds to a flux at the device-under-test (DUT) between 2E5 and 2E10 p/cm$^2$/s. The actual usable beam flux is somewhat constrained by limitations of the beam instrumentation currently in place. Online beam monitoring during an irradiation is done by a 120 mm plane-parallell transmission ionization chamber. The detector amplifies the beam current by a factor 130 and has been shown to have a linearity between 1E4 and 1E9 p/cm$^2$/s of better than 1%. Above 1E9 p/cm$^2$/s ion recombination increases rapidly which degrades the performance of the detector. A Faraday cup is used to calibrate the response of the transmission ionization chamber in units of fluence. A 60 mm diameter copper block collects the proton charge and is held in a vacuum tube, equipped with a HV guard ring to minimize electron contamination. The Faraday cup has been validated against a Faraday cup in Skandionkliniken, Uppsala, and were found to agree to better to 1%.

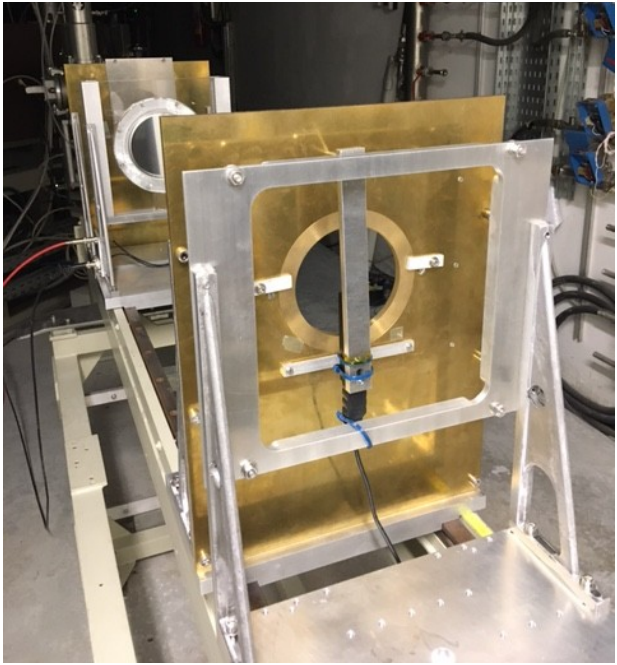

*Figure 2: The irradiation table with the DUT holder and the final 100 mm collimator in the foreground, and the beam instrumentation and beam-line in the background.*

A 100 mm diameter beam can thus be generated and monitored online in the flux interval 2E5 to 1E9 p/cm$^2$/s at a precision better than 5%.

*III.B Beam homogeneity*

A flat 100 mm beam at the DUT is achieved with an occluding double scattering system [1]. To validate that the homogeneity of the field is adequate for radiation tolerance tests, the beam radial profiles at 59 MeV (non-degraded), 40 MeV and 20 MeV were measured using gafchromic films. After being irradiated with ~10 Gy, the films were scanned and calibrated in units of dose. The radial beam profile for each energy was then derived by integrating the signal of each film in an annulus at a radius R from the center and having a thickness of 1 mm, for a radius between 0 and 80 mm. The result was normalized with the number of pixels in each circle. The resultant radial profiles for the three different beam energies are shown in Fig. 3. The error of each data point was computed as the standard deviation of the pixel intensity distribution in the corresponding annulus. The homogeneity of the primary beam is thus ±3%, estimated as the difference between the max and min value with respect to the average value for a radius <50 mm. The homogeneity for a 40 and 20 MeV beam is very similar, although the useful radius is somewhat reduced (~45 mm) as the lateral penumbra increases with decreasing energy. A 50 mm beam is however obtained within ±5% also at 20 MeV.

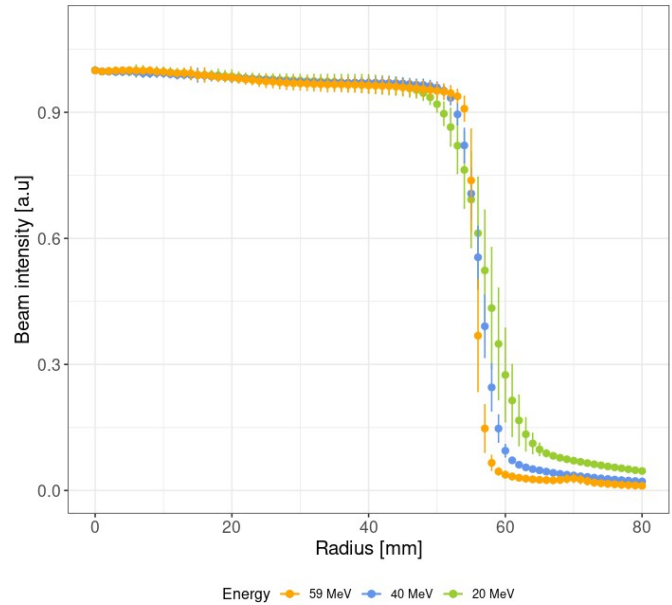

*Figure 3: The measured radial beam profiles at 59, 40 and 20 MeV. A homogeneity better than ±3% is achieved over the 100 mm diameter field at 59 MeV.*

*III.C Beam energy spread*

The (non-degraded) mean proton energy at the DUT has been estimated from measurements of the Bragg curve to (59.0 ± 0.5) MeV. The intrinsic beam energy spread was determined by minimizing the residuals between the measured and the simulated Bragg peak using the simulated energy spread as a free parameter. The Geant4 [5] package was used to simulate the beam-line. All beam parameters, except the energy spread, was set to their measured values. The quantity to minimize was the standard deviation of the distribution of the parameter q, defined as $q = (S - M)/M$, where S and M are the simulated and measured intensity vectors, and M the average measured intensity. An intrinsic energy spread of 0.3 MeV minimizes this quantity and gives a $\sigma(q)$ of 0.8%. When degrading the beam, the energy spread is however increasing which ultimately sets a lower limit on the usable energy range. The same procedure as above was used to estimate the energy spread at the DUT at each energy step between 20 and 59 MeV. At all three energies, $\sigma(q)$ was lower than 1.5%. Thus, the simulations well reproduce the degraded Bragg curve and therefore also the (degraded) beam energy distribution, and can be used to estimate the energy spread. The result is given in Fig. 4. Judging from these results, 20 MeV is a recommended lower energy limit of the beam-line.

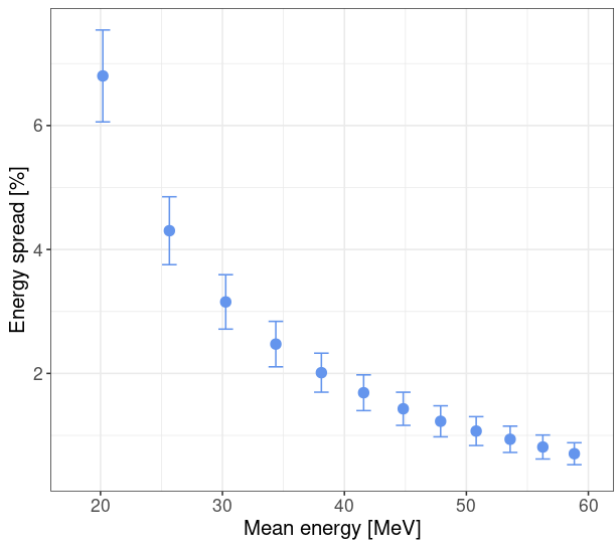

*Figure 4: The beam energy-spread versus mean energy, obtained from comparisons between measured and simulated bragg peaks.*

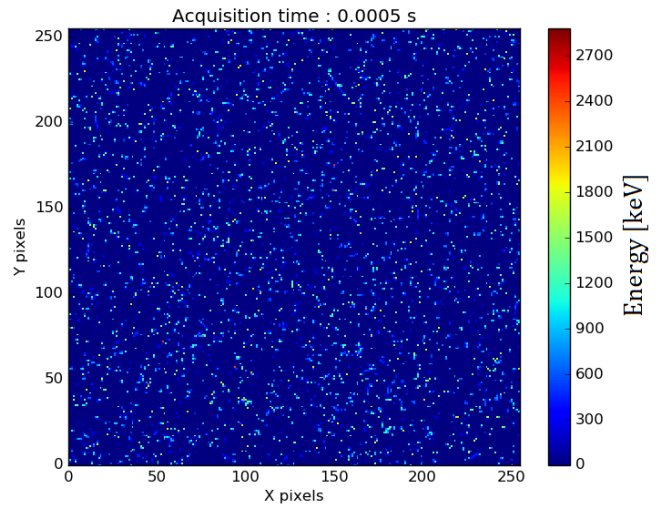

*Figure 5: The particle distribution measured at the center of the beam at 20 MeV.*

## IV  VALIDATION RESULTS

The CNES visited the CAL facility on the 28[th] of March 2022 and performed two experiments on the MEDICYC R&D beam-line to validate it for radiation hardness testing.

The first test consisted of measuring the deposited energy by the beam at various energies and compare the result with a measurement conducted at UCL (Université Catholique de Louvain) – an installation with very similar beam characteristics. The deposited energy distribution depends on the proton energy, energy distribution and contamination of secondary particles and thus provides a good test of similarity between the two beams. The measurements were done with a MiniPIX [6], a detector developed around the Timepix sensor [7]. Timepix is a hybrid semiconductor pixel detector developed by the Medipix collaboration at CERN, and is made up of a 500-µm thick silicon layer. Its 256 x 256 pixels are operated in time-over-threshold (ToT) mode to register the energy which is deposited by ionizing particles. The detector was placed in the center of the beam at the designated DUT position 4 cm behind the last collimator. Fig. 5 shows an example of the measured energy deposition over the field-of-view of the detector for a beam energy of 20 MeV. A comparison between the deposited energy distributions measured for the CAL and UCL beams is shown in Fig. 6 for a beam energy of 40 MeV. The histograms have been normalized to remove any potential fluence differences between the measurements. The two distributions agrees well and shows that the characteristics of the two beams are comparable.

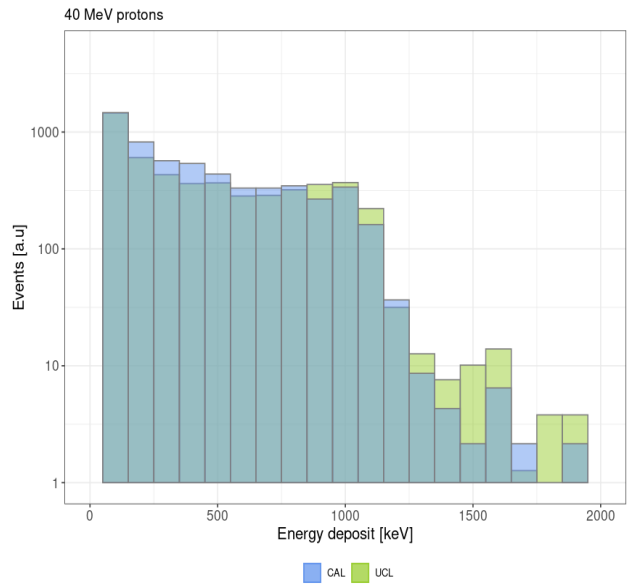

*Figure 6: The measured deposited energy distribution for the CAL and UCL beams at 40 MeV.*

The second experiment consisted of characterizing the Single Event Latch-up (SEL) response of a Brilliance Semiconductor BS62LV4006 SRAM memory by using the CNES TILU2 test-bench. The DUT was positioned in the beam using the fixation frame in Fig. 2. The test-bench was put behind the first concrete wall in the labyrinth in order to protect it as far as possible from secondary neutrons and gamma rays. The test-bench was supervised from the control room via an Ethernet connection. The tested DUT, CAL SN1, had previously been tested by CNES at the KVI-Center for Advanced Radiation Technology (KVI-CART), now Particle Therapy Research Center (PARTREC). It demonstrated a very high sensitivity to SEL under proton radiation. During the previous test-campaign at KVI two samples were tested having similar responses. At CAL-MEDICYC, only one device was characterized under a 20, 40 and 59 MeV proton beam at two different bias conditions, Vcc = 3.3 V and 5.0 V. Between 26 and 100 SELs were triggered in each irradiation. The resultant latch-up cross section is plotted in Fig. 7 together with the results from KVI overlaid. The results obtained at CAL are in agreement with the results obtained at KVI. Furthermore, it can be noted that the DUT is 3 to 5 times more sensitive with Vcc=5.0V, to be expected based on observations made on other BSI memories.

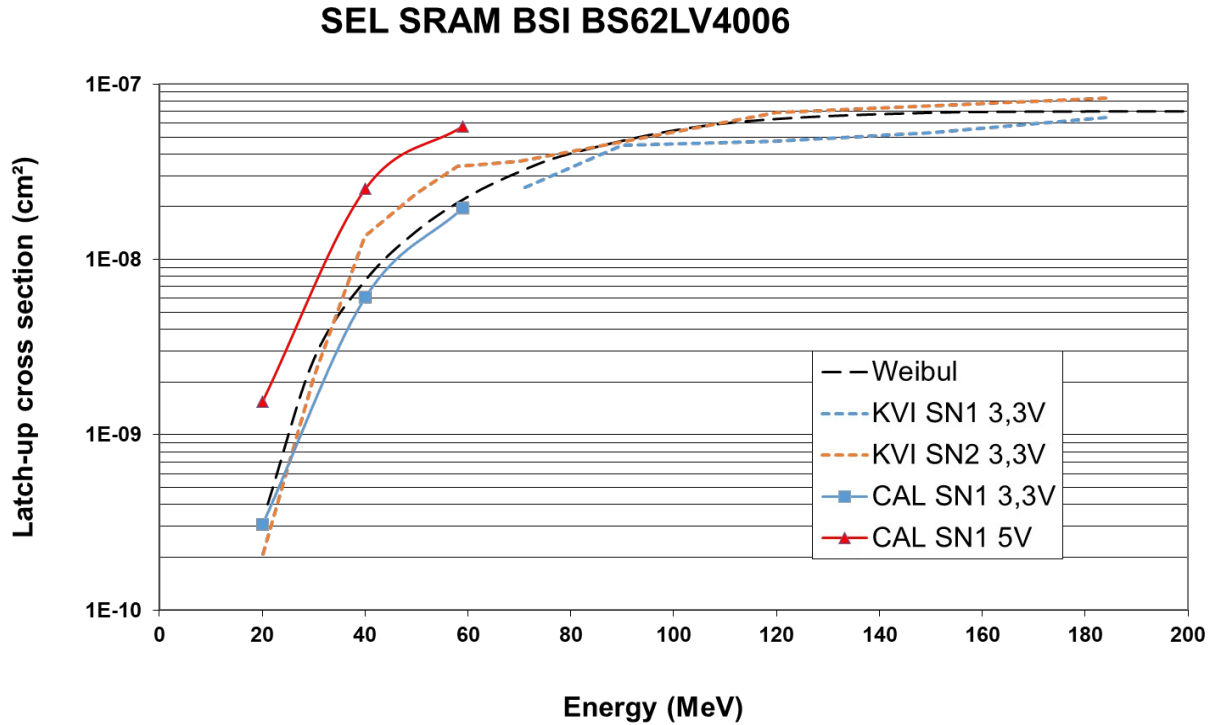

*Figure 7: The SEL cross section of a semiconductor BS62L4006 SRAM, measured at CAL and KVI.*

## V Conclusion

A common effort by the CAL and the CNES has been made to validate the MEDICYC R&D beam-line for radiation tolerance tests. The results presented in this proceeding demonstrate that the beam-line is well suited for this purpose. The available flux range covers the requirements for the vast majority of the types of tests targeted, at a flux measurement precision better than required. A beam homogeneity better than 3% is also significantly better than the requirements commonly demanded for radiation tolerance tests. A very low energy spread at 59 MeV has also been shown, although the spread increases somewhat at lower energies due to the passive energy selection method. Finally, the inter-comparison of the measured SEL response of a SRAM memory at two beam-lines with similar beam characteristics provides an additional cross-check of the complete beam-line setup at CAL.

Additional work is on-going at CAL to improve and stream-line radiation tolerance testing. A large diameter beam collector from Pyramid has been acquired which will improve the measurement precision especially at low flux and speed-up the beam calibration process. New transmission ionization chambers will also be installed to extend the measurement range to higher flux. Several improvements are also foreseen to improve the DUT positioning process, to be able to rapidly and precisely position the DUT at a given angle or distance from the beam.